\documentclass[11pt]{article}
\usepackage{moriond,epsfig}

\def\MyPhoto{\psfig{figure=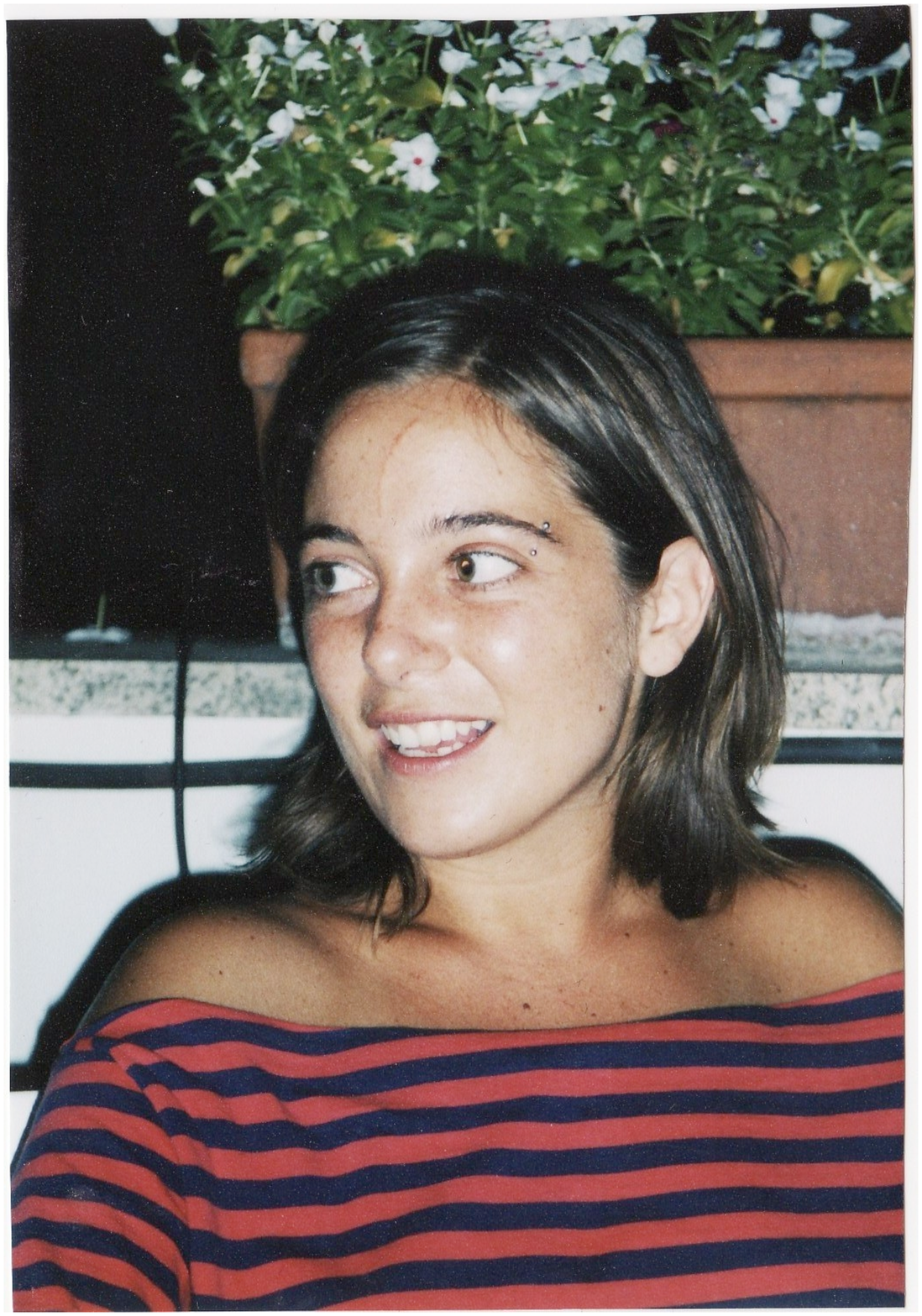,width=35mm}}

\def\be{\begin{equation}}
\def\ee{\end{equation}}
\def\bea{\begin{eqnarray}}
\def\eea{\end{eqnarray}}

\def\babar{BaBar~}
\def\blnu{$B^{+} \to l^{+} \nu_{l}$~}
\def\btaunu{$B^{+} \to \tau^{+} \nu_{l}$~}
\def\bmunu{$B^{+} \to \mu^{+} \nu_{l}$~}
\def\benu{$B^{+} \to e^{+} \nu_{l}$~}
\def\btag{$B_{tag}$~}
\def\bsig{$B_{sig}$~}
\def\br{${\cal B}$~}
\begin{document}
\vspace*{4cm}
\title{RARE DECAYS AT B FACTORIES}

\author{ELISABETTA BARACCHINI}

\address{Department of Physics and Astronomy, 4129 Frederick
Reines Hall, University of California, Irvine, CA 92697, USA\footnote{Formerly
at Universit\`a di Roma La Sapienza, Piazzale A. Moro 2, 00181 Rome, Italy}.}

\maketitle

\abstract{We will present a review of the most interesting results
on rare $B$ decays from the B Factories, based on the data collected by the 
\babar~\cite{Aubert:2001tu} and Belle~\cite{Bondar:1998gu} detectors at asymmetric $e^+ e^-$
colliders at the center of mass energy of the $\Upsilon(4S)$ resonance.}

\section{Introduction}
Rare $B$ decays have always been a standard probe for New Physics (NP)
searches as they
offer a prolific ground where to look for deviations
from the SM expectation. The very low SM rate of these decays often
make them unaccessible with the present experimental datasets, unless NP
effects enhance the rate up to the current experimental sensitivity.
In this view, if a suppressed decay is observed, clear sign of NP
can be claimed. On the other hand, if an upper limit (UL) is set, 
it can constraint NP scenarios.
B Factories provide an unique environment where to investigate these processes.
The high number of $B$ mesons pairs produced often allows to approach the needed
experimental sensitivity. Moreover, the clean environment and the closed 
kinematic of the initial state enable to obtain a very pure sample where
to look for these decays.

In this work we are going to present a review of results from both \babar
and Belle collaboration on the $B$ meson decays 
$B \to h^{(*)} \nu \bar{\nu}$, \blnu, $B^0 J/\psi \phi$ and
$B^{+} \to K^{-} \pi^+ \pi^-/K^+ K^- \pi^{-}$~\footnote{Charge
conjugation is implied through this paper, unless explicitly stated}.

\section{Analyses Overview}
The analyses presented in this work exploit different reconstruction techniques, 
depending on the particles present in the final state. In the decays
$B^0 J/\psi \phi$ and
$B^{+} \to K^{-} \pi^+ \pi^-/K^+ K^- \pi^{-}$ all particles
decaying from the signal $B$ are detectable, and thus a full
kinematic reconstruction of the event is possible.
On the other hand, \blnu and $B \to h^{(*)} \nu \bar{\nu}$ decays possess
one or more neutrinos in the final state, which clearly can not 
be detected. This particular characteristic calls for different
analysis techniques, which allows to deal with the lack of information
regarding these particles. Typically, the closed kinematic of an
$e^+e^-$ collision is exploited to constrain through energy and
four-vector conservation the $B \bar{B}$ pairs, after both particles
have been reconstructed. Different approaches can be employed in the selection
of the $B$ meson which is not decaying into the channel of interest (\btag):
a totally inclusive reconstruction is applied on the \btag, without trying to
identify its decay products, whenever the additional kinematic constraint
coming from the two-body nature of the signal $B$ (\bsig) can be
exploited, as in \bmunu and \benu analyses. The high efficiency obtainable
with this method has as drawback a poor energy resolution. On the other hand, 
when more than one neutrino is present in the event, a recoil technique is
needed:first, the \btag is reconstructed in either a semileptonic
$B_{sl} \to D^{(*)} l \nu$ or hadronic $B_{had} \to D Y$ ($Y=\pi, K$) system.
Then, the channel of interest is searched in the rest of the event (ROE), 
defined as the set of tracks and calorimeter deposits not associated with 
the \btag. The recoil method allows a very high resolution and purity, 
but has a low efficiency.

\section{Search for $b \to s \nu \bar{\nu}$ processes}

\subsection{Theoretical Introduction}

In the SM  $b \to s \nu \bar{\nu}$ transitions occurs through Flavour
Changing Neutral Current (FCNC) and are therefore forbidden at tree level.
As these processes occurs via one-loop box or electroweak penguin diagrams, 
they are expected to be highly suppressed. 
The expected branching ratios (\br) are
\br($B \to K* \nu \bar{\nu}$) = (6.8$^{+1.0}_{-1.1}$)$\times 10^{-6}$
and \br($B \to K \nu \bar{\nu}$) = (4.5$\pm 0.7$)$\times 10^{-6}$
\cite{Altmannshofer:2009ma}. However, this values can be enhanced in NP scenarios,
where several mechanisms can contribute to the rate. For example, in 
Ref.~\cite{Buchalla:2000sk}, non-standard $Z^0$ coupling can give rise
to an enhancement up to a factor 10. Moreover, new sources of missing
energy, such as light dark matter~\cite{Bird:2004ts} or unparticles
~\cite{Georgi:2007ek,Aliev:2007gr}, if accompanied by a $K^*$, would contribute
to the total rate. The kinematic of the decay can be described in terms of 
$s_{\nu \bar{\nu}} = m^2_{\nu \bar{\nu}}/m^2_B$, where $m_{\nu \bar{\nu}}$ is
the invariant mass of the neutrinos pair and $m_B$ is the $B$ meson mass.
As NP can strongly affect the decay in terms of $s_{\nu \bar{\nu}}$
shape~\cite{Altmannshofer:2009ma,Aliev:2007gr}, it is important to not rely on any
theoretical model when performing the analysis. 

\subsection{Analysis of $B \to h^{(*)} \nu \bar{\nu}$}

The Belle collaboration  analysis exploit the hadronic recoil
technique and looks for final states with 
$h^{(*)} = K^+, K^0_S, K^{*0}, K^{*+}, \pi^+, \pi^0, \rho^+, \rho^0, \phi $, 
using a 492 fb$^{-1}$ data sample~\cite{Chen:2007zk}. 
\btag candidates are reconstructed into a 
charged or neutral $D^{(*)}$ accompanied by
a charged $\pi$ or $\rho$ or $a_1$ or  $D^{(*)}_s$.
The $D^-$ mesons are reconstructed as $D^- \to K^0_S \pi^-,  K^0_S \pi^-
\pi^0,  K^0_S \pi^- \pi^+ \pi^-, K^+ \pi^- \pi^- $ and $K^+ \pi^-
\pi^- \pi^0$, while $ \bar{D}^{(*)0}$ as $ \bar{D}^{(*)0} \to K^+ \pi^-,
K^+\pi^- \pi^0, K^+ \pi^- \pi^+ \pi^-, K^0_S \pi^0, K^0_S \pi^+ \pi^-$,$K^+ K^-$ and 
$K^0_S \pi^- \pi^+ \pi^0$. The $D^{*-} (\bar{D}^{(*)0})$
mesons are reconstructed as $\bar{D}^{0} \pi^- (\bar{D}^{0} \pi^0$ and
$\bar{D}^{0} \gamma)$ and the $D^{*+}_s$ as $D^{*+}_s \to D^{+}_s \gamma$
and $D^{+}_s \to K^0_S K^+$ and $K^+ K^- \pi^+$.
No additional charged tracks or $\pi^0$ candidates are allowed in the
event and thus \bsig candidates are selected using the variable
$E_{\rm{ECL}} = E_{\rm{tot}}-E_{\rm{rec}}$, where $E_{\rm{tot}}$ and
$E_{\rm{rec}}$ are the total visible energy measured by the calorimeter
and the measured energy of reconstructed objects including the \btag
and the signal side $h^{(*)}$ candidate, respectively.

The dominant background comes from $BB$ decays involving a $b \to c$
transition. In order to suppress such decays, the momentum of the
$h^{(*)}$ candidate in the \bsig rest frame $P^{*}$ is required to
be $1.6$ GeV/c $< P^{*}<$ 2.5 GeV/c. The cosine of the missing momentum
in the lab frame is required to lie between -0.86 and 0.95 to rejects background
events where particles are missing along the beam pipe. 
Given the fact that reconstruction efficiencies for
the UL evaluation are estimated with MC simulation based on the SM, 
these two cuts introduce a SM dependence in the analysis results.
The background $E_{\rm{ECL}}$ distributions are normalized by the number
of events in the sideband region. None of the signal modes show a 
significant number of signal events. The observed number of events
$N_{obs}$, expected background $N_{b}$, reconstruction
efficiencies $\epsilon$ and the UL at 90$\%$ of confidence level obtained
with an extension of the Feldman-Cousins method
~\cite{Feldman:1997qc,Conrad:2002kn} are shown in Tab.~\ref{tab:hnunu_belle}.
\begin{table}
\caption{Observed number of events $N_{obs}$, expected background, $N_{b}$, reconstruction
efficiencies $\epsilon$ and the UL at 90$\%$ of confidence level for $B \to h^{(*)} \nu \bar{\nu}$}
 \begin{center}
 \begin{tabular}{ccccc} 
\hline \hline 
   Mode                       & $N_{obs}$  & $N_{b}$ & $\epsilon(\times 10^{-5})$ & UL \\
\hline \hline
    $K^{*0} \nu \bar{\nu}$       & 7 & 4.2 $\pm$ 1.4 & 5.1 $\pm$ 0.3 & $< 3.4 \times 10^{-4}$\\
    $K^{*+} \nu \bar{\nu}$       & 4 & 5.6 $\pm$ 1.8 & 5.8 $\pm$ 0.7 & $< 1.4 \times 10^{-4}$\\
    $K^{+} \nu \bar{\nu} $       &10 & 20.0 $\pm$ 4.0& 26.7 $\pm$ 2.9 &$ < 1.4 \times 10^{-5}$\\
    $K^{0} \nu \bar{\nu} $       &2 & 2.0 $\pm$ 0.9 & 5.0 $\pm$ 0.3 & $< 1.6 \times 10^{-4}$\\
    $\pi^+ \nu \bar{\nu} $      &33& 25.9 $\pm$ 3.9& 24.2 $\pm$ 2.6& $< 1.7 \times 10^{-4}$\\
    $\pi^0 \nu \bar{\nu} $      &11& 3.8 $\pm$ 1.3 & 12.8 $\pm$ 0.8 & $< 2.2 \times 10^{-4}$\\
    $\rho^0 \nu \bar{\nu}$      &21 & 11.5 $\pm$ 2.3 & 8.4 $\pm$ 0.5 & $< 4.4 \times 10^{-4}$\\
    $\rho^+ \nu \bar{\nu}$      & 15 & 17.8 $\pm$ 3.2 & 8.5 $\pm$ 1.1 & $<1.5 \times 10^{-4}$\\
    $\phi \nu \bar{\nu}  $    & 1 & 1.9 $\pm$ 0.9 & 9.6 $\pm$ 1.4 & $< 5.8 \times 10^{-5}$\\
\hline 
 \end{tabular}
 \end{center}
 \label{tab:hnunu_belle}
\end{table}

\subsection{Analysis of $B \to K^{*} \nu \bar{\nu}$}\label{subsec:hnunu_babar} 

The $B \to K^{*} \nu \bar{\nu}$  from \babar collaboration is performed in 
the recoil of both hadronic (HAD) and semileptonic (SL) system on a data
sample of 413 fb$^{-1}$~\cite{Aubert:2008fr}. 
The two
different tagging strategies provide non overlapping samples whose results
can be combined as independent measurements.
The event selection starts from the \btag reconstruction: in the SL analysis, 
neutral $D$ mesons are reconstructed in the $K^- \pi^+, K^- \pi^+ \pi^0,
K^- \pi^+ \pi^- \pi^+$ and $K^0_S \pi^+ \pi^-$ modes. Charged $D$ mesons 
are reconstructed in the $K^- \pi^+ \pi^+$ and $K^0_S \pi^+$ final states. 
In the HAD analysis, the $B_{\rm{had}}$ is reconstructed in 
$B_{\rm{had}} \to D Y$ where $Y = n\pi +mK + rK^0_S + q\pi^0$ with
$n +m  + r +q <$ 6 and $D$ is a generic charmed meson. About 1000 different
decay chains are considered. Charmed mesons are reconstructed in the same final 
states used in the SL analysis, along with the additional channels
$D^+ \to K^+ \pi^- \pi^+ \pi^0, K^0_S \pi^+ \pi^- \pi^+, K^0_S \pi^+ \pi^0$.
For each reconstructed tagging $B$, a $K^*$ is searched in the ROE
and reconstructed in the $K^+\pi^-$, $K^0_S \pi^+$ or $K^+ \pi^0$ mode.
Considering that signal events have no additional neutral particles produced
in association with the $K^*$, one of the most discriminating variable between
signal and background is, as in the Belle analysis, the extra neutral energy
$E_{\rm{extra}}$, the sum of the energies of the calorimeter neutral clusters
not used to reconstruct either the \bsig or the \btag.
In the SL analysis, the signal yield is extracted through a Maximum 
Likelihood (ML) fit to the final $E_{\rm{extra}}$ distribution, after selection
criteria are applied to suppress the continuum background. In the HAD analysis,
a loose selection is applied and all discriminants variables (including
$E_{\rm{extra}}$) are used as inputs for a Neural Network (NN),
whose output variable $NN_{\rm{out}}$ is fitted to extract the number of
signal events. No significant signal is observed in the two analysis and a 
Bayesian approach is employed to set the UL at 90$\%$ of confidence level.
The UL from the SL analysis are ${\cal B}(B^0 \to K^{*0} \nu \bar{\nu}) <$ 18 
$\times 10^{-5}$ and ${\cal B}(B^+ \to K^{*+} \nu \bar{\nu})<$  9 
$\times 10^{-5}$, the UL from the HAD analysis are 
${\cal B}(B^0 \to K^{*0} \nu \bar{\nu}) <$  11 $\times 10^{-5}$ and
${\cal B}(B^+ \to K^{*+} \nu \bar{\nu}) <$  21 $\times 10^{-5}$. The combined UL
are ${\cal B}(B^0 \to K^{*0} \nu \bar{\nu}) <$  12 $\times 10^{-5}$, 
${\cal B}(B^+ \to K^{*+} \nu \bar{\nu})<$  8 $\times 10^{-5}$ and
${\cal B}(B \to K^{*} \nu \bar{\nu})<$  8 $\times 10^{-5}$.

These results are at the moment the most restrictive UL on these
decay channels and are the first completely model independent measurement.

\section{Search for $B^+ \to l^+ \nu$ processes}

\subsection{Theoretical Introduction}

In the SM the purely leptonic $B$ decays $B^+ \to l^+ \nu$ ($l= e, \mu, \tau$)
proceed through the annihilation of the two quarks in the meson to form a 
virtual $W$ boson. The branching ratio can be cleanly calculated in the SM
%
%
and is sensitive to the Cabibbo Kobayashi Maskawa matrix element $V_{ub}$ and the
$B$ decay constant $f_B$, which describes the overlap of the quark wave 
functions within the meson and which is currently the major
source of uncertainty in the {\cal B} calculation.
Assuming $\tau_B$ = 1.638 $\pm$ 0.011 ps, $V_{ub}$=(4.39 $\pm$ 0.33)
$\times 10^{-3}$ determined from inclusive charmless semileptonic $B$ 
decays~\cite{Barberio:2006bi} and 
$f_B$ = 216$ \pm $22 MeV from lattice QCD calculation~\cite{Gray:2005ad},
the SM estimate of ${\cal B}$(\btaunu) is $(1.59 \pm 0.40)\times 10^{-4}$ .
Due to helicity suppression, \bmunu and \benu are suppressed by 
factors $m_{\mu,e}^2/m_{\tau}^2$ with respect to \btaunu, leading to expected   
branching fractions of ${\cal B}$(\bmunu) = $(5.6 \pm 0.4) \times 10^{-7}$ and
${\cal B}$(\benu) = $(1.3 \pm 0.4) \times 10^{-11}$.
Purely leptonic $B$ decays are sensitive to physics beyond the SM,
where additional heavy virtual particles contribute to the annihilation processes.
Charged Higgs boson effects may greatly enhance or suppress the 
branching fraction in some two-Higgs-doublet models~\cite{Hou:1992sy}.
Moreover, in a SUSY scenario at large $\tan \beta$, non-standard effects in helicity-suppressed 
charged current interactions are potentially observable, being strongly $\tan \beta$-dependent
and leading to~\cite{Hou:1992sy}:
$\frac{{\cal B}(B^{+} \rightarrow l^{+} \nu_{l})_{\rm{exp}}}{{\cal B}(B^{+} \rightarrow l^{+} \nu_{l})_{\rm{SM}} } \approx ( 1-  \tan^2 \beta \frac{m_B^2}{M_H^2} )^2 $.
These decays are also potential probes for Lepton Flavour Violation (LFV)
in the ratios $R^{\mu/\tau}_B = {\cal B}$(\bmunu)/${\cal B}$(\btaunu) and
$R^{e/\tau}_B = {\cal B}$(\bmunu)/${\cal B}$(\btaunu)~\cite{Isidori:2006pk}.

\subsection{Analysis of \blnu in the SL Recoil}
The \blnu analysis from the \babar collaboration is performed
in the recoil of a semileptonic system on a data sample
of 418 fb$^{-1}$~\cite{Aubert:2008gx}. The \btag is reconstructed
in the set of semileptonic $B$ decay mode $B^- \to D^0 l^- \bar{\nu} X$,
where $l = e,\mu$ and $X$ can be either nothing or a transition
particle from a higher mass charm state decay, which is not reconstructed
(although tags consistent with a neutral $B$ decay are vetoed). The
$D^0$ is reconstructed in the same modes as in~\ref{subsec:hnunu_babar}.
Since $\tau$ decays before reaching active detector elements, the 
\btaunu signal is searched for in both leptonic and hadronic $\tau$
decay modes: $\tau^+ \to e^+ \nu_e \bar{\nu}_{\tau},
\mu^+ \nu_{\mu} \bar{\nu}_{\tau} , \pi^+ \bar{\nu}_{\tau}$ and 
$\pi^+ \pi^0 \bar{\nu}_{\tau}$. The lepton momentum in the 
\bsig rest frame is used to separate electrons and muons from
\bmunu or \benu and from $\tau$ decays. Backgrounds consists primarily
of $B^+ B^-$ events in which the \btag has been correctly reconstructed
and the recoil side contains one signal candidate track and additional
particles which are not reconstructed by the tracking detectors
or calorimeter. Typically, these events contain $K^0_L$ candidates
and/or neutrinos. In addition, some excess events in data, most likely
from two-photon and QED processes which are not modeled in the MC
simulation, are also seen. Multiple variables are used to suppress
backgrounds and are combined in two likelihood ratios (LHRs), which 
are probability distributions designed to produce maximum separation between
signal and background. Among them, two of the most powerful are
the ratio of the second and the zeroth Fox-Wolfram moment 
$R2$~\cite{Fox:1978vu}  and the cosine of the angle between the
$B$ meson momenta and the $D^0 l $ candidate in the $\Upsilon(4S)$ frame.
Also in this analysis, the total energy recorded in the detector and not
assigned to either the \bsig or the \btag $E_{\rm{extra}}$ is used to
select signal decays.
The number of expected background events is estimated from the 
$E_{rm{extra}}$ sidebands.
The observed number of events
$N_{obs}$, expected background $N_{b}$, reconstruction
efficiencies $\epsilon$ and the branching ratio ${\cal B}$ or UL at
 90$\%$ of confidence level obtained with the Feldman-Cousins method
~\cite{Feldman:1997qc} are shown in Tab.~\ref{tab:blnu_babar}.

\begin{table}
\caption{Observed number of events $N_{obs}$, expected background $N_{b}$, reconstruction
efficiencies $\epsilon$ and ${\cal B}$ or UL at 90$\%$ of confidence level for \blnu}
 \begin{center}
 \begin{tabular}{ccccc} 
\hline \hline 
   Mode         & $N_{obs}$  & $N_{b}$ & $\epsilon(\times 10^{-4})$ & ${\cal B}$ \\
\hline \hline
$\tau^+ \nu$      & 610 & 521 $\pm$ 31  & 10.54 $\pm$ 0.41 & (1.8 $\pm$ 0.8 $\pm$ 0.1)$\times  10^{-4}$\\
$\mu^+ \nu $     &11 & 15 $\pm$ 10 & 27.1 $\pm$ 1.2 & $ <11 \times 10^{-6}$ @ 90$\%$ CL\\
$e^+ \nu $ & 17 & 24 $\pm$ 11 & 36.9 $\pm$ 1.5 &  $< 7.7 \times 10^{-6}$ @ 90$\%$ CL\\
\hline 
 \end{tabular}
 \end{center}
 \label{tab:blnu_babar}
\end{table}
Combined with the previous measurement in the HAD recoil
~\cite{Aubert:2007xj}, the
\btaunu average from \babar is ${\cal B} (B^+ \to \tau^+ \nu) = (1.8 \pm 0.6 )
\times 10^{-4}$. 

\subsection{Inclusive Analyses of \blnu $(l=e,\mu)$ }
\blnu are two-body decays, so the lepton is produced mono-energetic in the
\bsig rest frame. Thus, in inclusive analyses, the highest momentum lepton in the event
is searched and assign to the signal side, and all other charged
tracks and neutral clusters in the event are used to reconstruct the
\btag, without trying to identify the direct decay products of the
\btag. 
The momentum direction of the reconstructed \btag is used to
boost the lepton candidate in the \bsig rest frame and thus refine
the estimate of the lepton momentum in this frame ($p^*$).
The two most significant backgrounds are $B$ semileptonic decays
involving $B \to X_{u,c} l \nu $ transitions where the endpoint of the
lepton spectrum approach that of the signal, and non-resonant $q \bar{q}$
events.  

Belle analysis~\cite{Satoyama:2006xn} is on a data sample
of 253 fb$^{-1}$ and applies tight 
cuts on lepton momentum in both CM
frame $(p_{\rm{CM}})$ and \bsig rest frame $p^*$ to remove 
 $B \to X_{u,c} l \nu $ backgrounds, and exploit the combination of
 modified Fox-Wolfram moments~\cite{Fox:1978vu,Abe:2001hk} in a Fisher
discriminant to suppress continuum events. A cut on the \btag
$\Delta E = E_B - E_{\rm{beam}}$, where $E_B$ is the reconstructed
energy of the \btag and $E_{\rm{beam}}$ the beam energy in the CM frame,
 is applied to refine the selection. The final selection
efficiency is $(2.2 \pm 0.1)\%$ for \bmunu and $(2.4 \pm 0.1)\%$
for \benu.
The yields are extracted from a ML fit to the beam-energy constrained mass
$M_{bc} = \sqrt{E^2_{\rm{beam}}-p^2_B}$ distributions, where $p_B$ is the reconstructed
momentum of the \btag, 
and 4.1 $\pm$ 3.1 signal events are observed for the muon mode and
-1.8 $\pm$ 3.3 for the electron mode.
The 90$\%$ confidence level for the upper limit on the branching fraction
${\cal B}_{90}$ are defined by $0.9 = \int^{{\cal B}_{90}}_{0} {\cal L}({\cal B})
d{\cal B} / \int^{\infty}_{0} {\cal L}({\cal B})
d{\cal B}$ and are ${\cal B}$ (\bmunu)$< 1.7 \times 10^{-6}$ and
 ${\cal B}$ (\benu) $< 9.8 \times 10^{-7}$ 

The 90$\%$ confidence level UL for the electron mode is currently the
most stringent measurement available.

\babar analysis~\cite{Aubert:2009rk} uses an integrated luminosity
of 426 fb$^{-1}$ and choose to
combine five different topological and kinematical variables,
optimized separately for each mode, in a Fisher discriminant for 
$q \bar{q}$ background suppression. A cut on the \btag $\Delta E$
is applied for the muon mode in order to remove continuum background,
while two linear combination of \btag $\Delta E$ and \btag $p_{T}$ are used
for the electron mode in order to reject also the background coming from two-photon
events. The final selection efficiencies are $(6.1 \pm 0.2)\%$
for the muon mode and $(4.7 \pm 0.3)\%$ for the electron mode.
The two-body nature of the decay is exploited by
combining $p^*$ and $p_{\rm{CM}}$ in a second Fisher discriminant, whose
output $p_{\rm{FIT}}$ is used in combination with the \btag 
$m_{ES} = \sqrt{E_{\rm beam}^{2}-|\vec{p}_B|^{\;2}}$
a ML fit to extract the yields. The number of observed signal events
is $1.4 \pm 17.2$ for \bmunu and $17.9 \pm 17.6$ for \benu. The 
 90$\%$ confidence level ULs are evaluated with a Bayesian approach
and are ${\cal B}$ (\bmunu)$< 1.0 \times 10^{-6}$ and
 ${\cal B}$ (\benu)$< 1.9 \times 10^{-6}$ 

The 90$\%$ confidence level UL for the muon mode is more restrictive
than any previous measurements.

\section{Search for $B^0 \to J/\psi \phi$ decay}

\subsection{Theoretical Introduction}
Studies of exclusive $B$ meson decays to charmonium play an important role in
exploring $CP$ violation~\cite{Aubert:2001nu,Abe:2001xe} and establish 
the Cabibbo-Kobayashi-Maskawa picture.
The decay $B^0 \to J/\psi \phi $ is expected to proceed mainly via a 
Cabibbo-suppressed and a color-suppressed transition $(b \to c \bar{c} d)$
with rescattering. In $B$ decays, effects presumably due to rescattering
have been seen in various decay processes, as $B^0 \to D^-_s K^+$
~\cite{Aubert:2006hu,Krokovny:2002pe}, and they may play an important role 
in understanding patterns of $CP$ asymmetries in $B$ decays to two charmless 
pseudoscalars~\cite{Chua:2002wk}.

\subsection{Analysis of $B^0 \to J/\psi \phi $}
The Belle analysis of $B^0 \to J/\psi \phi $ decay is performed on a 
data sample of 605 fb$^{-1}$~\cite{Liu:2008bt}. 
Candidates for $B^0 \to J/\psi \phi$ decays are reconstructed from the decay
$J/\psi \to l^+ l^- (l= e, \mu)$ and $\phi \to K^+ K^-$. 
$B^0$ are selected through the usual $M_{bc}$ and $\Delta E$ variables
and the signal yield is extracted through a ML fit to the $\Delta E$
distribution. 

The dominant background comes from $B \bar{B}$ events with a $B$ decays
to a $J/\psi$, in particular $B^0 \to J/\psi K^{*0}(892) [\to K^- \pi^+]$
and $B^{0/-} \to J/\psi K_1(1270)[\to K^- \pi^+ \pi^{0/-}]$. In both cases, 
a pion is misidentified as a kaon, and in the latter case the other pion
is missed. The former has a peak at $\Delta E \sim 0.1$ GeV, while the
latter has a broad peak in the negative $\Delta E$ region. These background
are taken into account into the ML fit, as well as the remaining not-peaking
combinatorial background. 
The number of observed signal events is 4.6$^{+3.1}_{-2.5}$ with a significance
of 2.3 $\sigma$ and the
 90$\%$ confidence UL, extracted by a 
frequentistic method using ensembles of pseudo-experiments,
is ${\cal B}(B^0 \to J/\psi \phi)<$ 9.4 $\times 10^{-7}$. The UL is the 
most restrictive up to date and improve of about one order of magnitude
the previous result~\cite{Aubert:2003ii}.

\section{Search for $b \to qq \bar{d}/qq\bar{s}$ Processes}

\subsection{Theoretical introduction}

$b \to qq \bar{d}/\bar{s}$ transitions are highly suppressed in the SM:
compared with the penguin (loop) transition $b \to q \bar{q} d/s$, they
are additionally suppressed by the small Cabibbo-Kobayashi-Maskawa matrix
element factor $|V_{td} V_{ts}^{*}| \simeq 3 \times 10^{-4}$, leading to
a predicted branching fractions of only ${\cal O}(10^{-14})$ and
${\cal O}(10^{-11})$, respectively~\cite{Huitu:1998vn,Fajfer:2006av}. These branching ratios can be
significantly enhanced in SM extensions as the minimal supersymmetric 
standard model (MSSM) with or without conserved $R$ parity, or models
containing extra $U(1)$ gauge bosons~\cite{Fajfer:2006av}. Observation of the
decays $B^- \to K^+ \pi^- \pi^-$ and $B^- \to K^- K^- \pi^+$ would be
clear experimental signature for $b \to dd \bar{s}$ and $b \to ss \bar{d}$
quark transitions, which have already been searched in these and other
decay modes without success yet.

\subsection{Analysis of $B^- \to K^+ \pi^- \pi^-/K^- K^- \pi^+$}
The \babar $B^- \to K^+ \pi^- \pi^-/K^- K^- \pi^+$ analysis uses
426 fb$^{-1}$ of integrated luminosity~\cite{Aubert:2008rr}.
$B^- \to K^+ \pi^- \pi^-(K^- K^- \pi^+)$ candidates are selected by
combining a charged kaon (pion) candidate with two charged pion (kaon),
each of which has charge opposite to the kaon (pion).
In order to avoid potentially large source of background arising from $B$
decays mediated by the favored $b \to c$ transitions, 
$B$ decays in which the pairs of daughter tracks have 
invariant mass combinations in the range 1.76 $<m_{K \pi}<$ 1.94 GeV/c$^2$,
2.85   $<m_{K \pi}<$  3.25 GeV/c$^2$ and 3.65  $<m_{K \pi}<$  3.75 GeV/c$^2$.
These vetoes remove events containing the decays $D^0 \to K^- \pi^+$, 
$J/\psi \to l^+ l^-$ and $\phi(2S) \to l^+ l^-$, respectively,
 where the leptons in the $J/\psi$ and $\phi(2S)$ decays are misidentified
as pions and kaons. The final selection efficiencies are
21.6$\%$ for $B^- \to K^- \pi^+ \pi^+$ and 17.8$\%$ for 
$B^- \to K^- K^- \pi^+$.
Continuum events represent the dominant background: in order to discriminate
against it, five variables, comprising topological quantities as well
as the flavour properties of the recoiling $B$ meson and
the proper time difference between the two $B$s, are combined into a NN. 
Residual backgrounds, arising from decays topologically similar to the
signal but with some misreconstruction, can be divided in five categories, depending
on their shape in $m_{ES}$ and $\Delta E$, and are taken into account
in the fit to the yield. 
The number of signal events is extracted through a ML fit to $m_{ES}$,
$\Delta E$ and the NN output $NN_{out}$ and is found to be
22 $\pm$ 43 for $B^- \to K^- \pi^+ \pi^+$ and -26 $\pm$ 19 for
$B^- \to K^- K^- \pi^+$.
A frequentistic Feldman-Cousins method is employed to
obtain the 90$\%$ confidence level UL, which are 
 ${\cal B}(B^- \to K^- \pi^+ \pi^+) <$ 7.4 $\times 10^{-7}$
and  ${\cal B}(B^- \to K^- K^- \pi^+)< $ 4.2 $\times 10^{-7}$
and improve of about a factor 3 previous measurements~\cite{Aubert:2003xz}.

\section{Conclusions}
In this work we have presented a review of the most interesting results on
rare $B$ decays at the B Factories from \babar and Belle experiments. 

Rare decays posses high interest, as they are standard probe for NP
searches, given the low decay rate expected in the SM. Their study is
complementary to the direct exploration of the energy frontier and in some
cases can access even higher energy scales. We have seen how the improved 
analysis techniques and the huge integrated luminosity from both \babar
and Belle experiments allow today to reach ${\cal O}(10^{-6}-10^{-7})$
sensitivity and how, even if only UL, the results on these decay are already
able to pose interesting constraints on various NP scenarios.
Nonetheless, decays with undetectable particles in the final state
will not be measurable at the LHC and a Super Flavour Factory will
be needed in order to obtain improved measurements.

\end{document}